\documentclass[aps,prl,twocolumn,amsmath,amssymb,superscriptaddress,longbibliography]{revtex4-2}
\usepackage[english]{babel}
\usepackage{xcolor}
\usepackage{amssymb} 
\usepackage{dcolumn}
\usepackage{bm}
\usepackage{graphicx}
\usepackage{amsmath}
\usepackage{braket}

\usepackage{CJKutf8}

\usepackage{graphicx}        

\graphicspath{{pict/}{}}

\usepackage[normalem]{ulem}

\usepackage{dcolumn}
\usepackage{bm}
\usepackage[pdfstartview=FitH, CJKbookmarks=true, bookmarksnumbered=true, bookmarksopen=true, colorlinks=true, pdfborder=001, citecolor=blue, linkcolor=blue, urlcolor=blue, linktocpage=true] {hyperref}

\makeatletter

\begin{document}

\title{Quantum Interference Amplifies Weak Chirality into Giant Quantum Nonreciprocity}

\author{Jing Tang}
\affiliation{School of Physics and Optoelectronic Engineering, Guangdong University of Technology, Guangzhou 510006, China}
\affiliation{Guangdong Provincial Key Laboratory of Sensing Physics and System Integration Applications, Guangdong University of Technology, Guangzhou, 510006, China}

\author{Yuangang Deng}
\email{dengyg3@mail.sysu.edu.cn}
\affiliation{Guangdong Provincial Key Laboratory of Quantum Metrology and Sensing $\&$ School of Physics and Astronomy, Sun Yat-Sen University, Zhuhai 519082, China}

\date{\today}
\begin{abstract}
Quantum nonreciprocity at few-photon level typically requires strong symmetry breaking, posing significant experimental challenges. Here we demonstrate that phase-controlled quantum interference can amplify weak chirality into giant quantum nonreciprocity. We consider two phase-programmable atoms coupled to a spinning whispering-gallery-mode resonator, where interference dramatically amplifies the effect of weak Fizeau splitting. This mechanism generates pronounced directional asymmetry in photon statistics, featuring bright antibunched emission in one direction and strongly bunched emission in the opposite direction. Remarkably, both correlation and brightness isolations obey phase-controlled power-law scaling with Fizeau splitting, reaching up to 65~dB and 17.3~dB, respectively. Our results establish interference-enhanced weak chirality as a powerful route toward directional nonclassical light sources.
\end{abstract}

\date{\today}  

\maketitle

{\em Introduction}.---Optical nonreciprocity~\cite{shoji2014magneto,sounas2017non,shen2016experimental,PhysRevLett.118.033901,gu2017microwave}, directional signal transport, is a key resource for photonic information processing, enabling optical isolation~\cite{sollner2015deterministic,dong2021all,jalas2013and,sayrin2015nanophotonic}, circulation~\cite{PhysRevLett.121.203602,ruesink2018optical,PhysRevA.91.053854}, and sensors~\cite{fleury2015invisible,yang2015invisible}. Extending nonreciprocity to quantum regime is crucial for few-photon devices, quantum networks~\cite{kimble2008quantum,hu2019coherent,lodahl2017chiral}, quantum routers~\cite{shomroni2014all,PhysRevLett.120.060601}, and nonclassical light sources~\cite{PhysRevLett.123.233604,PhysRevLett.121.153601,PhysRevLett.133.043601}. Existing approaches based on magneto-optical effects~\cite{shi2015limitations,hu2021noiseless}, chiral light matter coupling~\cite{scheucher2016quantum,fan2012all,zhang2025chirality}, parametric amplification~\cite{PhysRevLett.128.083604,PhysRevX.7.031001,PhysRevLett.120.023601}, and non-Hermitian engineering~\cite{chang2014parity,del2022non,cao2020reservoir,song2024experimental}, generally require strong symmetry breaking, such as large magnetic bias~\cite{guo2018significant,PhysRevLett.101.113902}, strong chiral coupling~\cite{lodahl2017chiral,PRXQuantum.6.020101}, or gain-loss asymmetry near exceptional points~\cite{hodaei2017enhanced,el2018non,peng2014parity}, substantially limiting scalability and experimental implementations. Achieving strong quantum nonreciprocity from weak chirality therefore remains a central challenge.  

Spinning whispering-gallery-mode (WGM) resonators provide a magnetic-free platform for nonreciprocity~\cite{PhysRevLett.125.123901,PhysRevLett.102.213903,estep2014magnetic}. Rotation lifts the degeneracy between clockwise (CW) and counterclockwise (CCW) modes via Sagnac-Fizeau effect~\cite{malykin2000sagnac,Nature558569}, enabling directional transport~\cite{fleury2014sound,popa2014non,PhysRevLett.120.125501}, entanglement~\cite{PhysRevLett.125.143605,yang2020nonreciprocal,jiao2022nonreciprocal,PhysRevA.95.063807}, photon blockade~\cite{graf2022nonreciprocity,PhysRevA.101.013826,PhysRevA.100.053832}, and quantum amplification~\cite{PhysRevX.5.021025,PhysRevLett.120.023601}. In realistic devices, however, Fizeau splitting is intrinsically weak and constrained by mechanical stability and rotation speed~\cite{PhysRevLett.114.053903,Liang:17}. This raises a key question: can weak chiral symmetry breaking be amplified into strong quantum nonreciprocity, producing directional antibunched and strongly bunched photon emission with high fidelity?

\begin{figure}[ht]
\includegraphics[width=0.9\columnwidth]{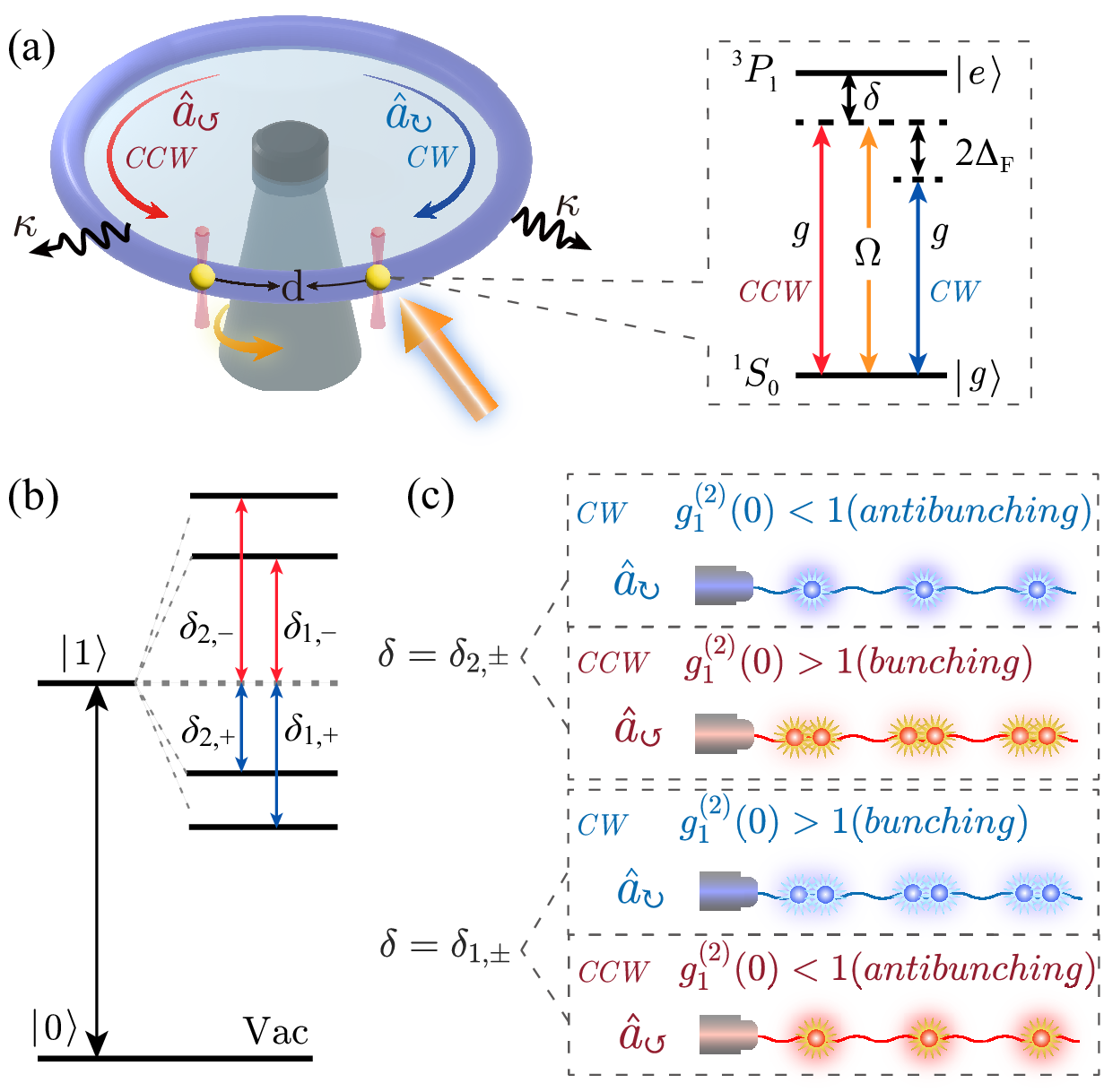}
\caption{(a) Schematic of interference-enhanced amplification of weak chirality in two atoms coupled to a spinning WGM resonator. CW and CCW modes couple dipole-forbidden transition between $^1S_0=|g\rangle$  and $^3P_1=|e\rangle$. (b) Anharmonic energy spectrum for $\phi=\pi/2$ and $\Delta_F>0$, showing four vacuum-Rabi branches $\delta_{1,\pm}$ and $\delta_{2,\pm}$. (c) Weak chiral asymmetry is amplified into giant directional asymmetry in photon correlations and emission brightness through quantum interference.
}
\label{model}
\end{figure} 

In this Letter, we show that phase-controlled quantum interference provides a general mechanism for amplifying weak chirality into giant quantum nonreciprocity. We consider two phase-programmable atoms coupled to a spinning WGM resonator, where the relative atomic phase controls interference between excitation pathways. In nonspinning limit, interference alone enables continuous crossover from antibunched single-photon emission to strongly bunched two-photon bundles. Introducing a finite Fizeau splitting renders system highly sensitive to weak mode asymmetry, producing pronounced direction-dependent photon statistics. This results in a clear separation between antibunched and bunched emission along opposite propagation directions. Both correlation and brightness isolations exhibit phase-controlled power-law scaling with Fizeau splitting, revealing an interference-enhanced amplification mechanism beyond conventional nonreciprocal schemes. Our results establish weak-chirality amplification as a promising route toward programmable chiral networks~\cite{kimble2008quantum,hu2019coherent,lodahl2017chiral} and directional nonclassical states~\cite{PhysRevLett.123.233604,PhysRevLett.121.153601,PhysRevLett.133.043601,PhysRevLett.125.143605,yang2020nonreciprocal,jiao2022nonreciprocal,PhysRevA.95.063807}. Such an interference-enhanced  mechanism may further provide a sensitive platform for chiral-molecule detection~\cite{bougas2022absolute,patterson2013enantiomer}, where weak optical-activity-induced shifts translate into large asymmetries in photon correlations.

{\em WGM resonator-coupled atomic array}.---We consider a minimal cavity-QED model of two phase-programmable atoms coupled to spinning WGM resonator~\cite{PhysRevLett.126.233602,aoki2006observation,PhysRevLett.110.213604,shomroni2014all}, as sketched in Fig.~\ref{model}(a). The  CW and CCW modes couple to the atomic transition $|g\rangle\!\leftrightarrow\!|e\rangle$ with equal strength $g$. The relative atomic positions imprint a tunable phase $\phi=2\pi d/\lambda$, where $d$ is interatomic separation and $\lambda$ is cavity wavelength. For typical tweezer separations ($d\sim 5\,\mu$m) much smaller than resonator radius ($\sim 500\,\mu$m), both atoms couple efficiently to WGM modes. CCW rotation lifts CW-CCW degeneracy via Fizeau splitting $\Delta_F$~\cite{malykin2000sagnac}. In the rotating frame of the pump, the system Hamiltonian reads
\begin{align}\label{Ham}
{\cal \hat H}
&=(\Delta_c+2\Delta_F)\hat a_{\circlearrowright}^\dag \hat a_{\circlearrowright}
+\Delta_c \hat a_{\circlearrowleft}^\dag \hat a_{\circlearrowleft}
+\sum_{i=1}^{2}\left(\frac{\delta}{2}\hat \sigma_i^z+\Omega \hat \sigma_i^x\right)
\nonumber\\
&+ g(\hat a_{\circlearrowright}^\dag+\hat a_{\circlearrowleft}^\dag)\hat \sigma_1^-
+g(e^{i\phi}\hat a_{\circlearrowright}^\dag+e^{-i\phi}\hat a_{\circlearrowleft}^\dag)\hat \sigma_2^- + {\rm H.c.},
\end{align}
where $\hat a_{\circlearrowright}$ ($\hat a_{\circlearrowleft}$) denotes annihilation operator of CW (CCW) mode, $\hat \sigma_i^{x,y,z}$ are Pauli operators for $i$th atom with $\hat \sigma_i^\pm=(\hat \sigma_i^x\pm i\hat \sigma_i^y)/2$, $\delta$ is atom-pump detuning, $\Delta_c$ is cavity-pump detuning referenced to CCW mode, and $\Omega$ denotes Rabi coupling of transverse pump. We focus on cavity-atom resonance regime $\Delta_c=2\delta$.

The essential physics arises from the interplay of phase-controlled interference and weak chiral symmetry breaking. For $\Delta_F=0$, CW and CCW modes remain degenerate, and phase $\phi$ only redistributes excitation amplitudes among collective pathways, preserving reciprocity in both photon number and statistics. A finite $\Delta_F$ breaks this degeneracy, rendering the corresponding pathways spectrally inequivalent. This effect already appears in single-excitation manifold, which splits into four branches [Fig.~\ref{model}(b)]. For $\phi=\pi/2$, the single-photon resonances are located at $\delta_{1,\pm}/g=\pm1$ and $\delta_{2,\pm} = (-\Delta_F \pm \sqrt{\Delta_F^2 + 16 g^2})/4$. The resulting asymmetry $\delta_{2,+} \neq |\delta_{2,-}|$ directly reflects rotation-induced chirality, with quantum interference strongly enhancing sensitivity, leading to giant nonreciprocity in both photon brightness and correlations. 

\begin{figure}[ht]
\includegraphics[width=0.95\columnwidth]{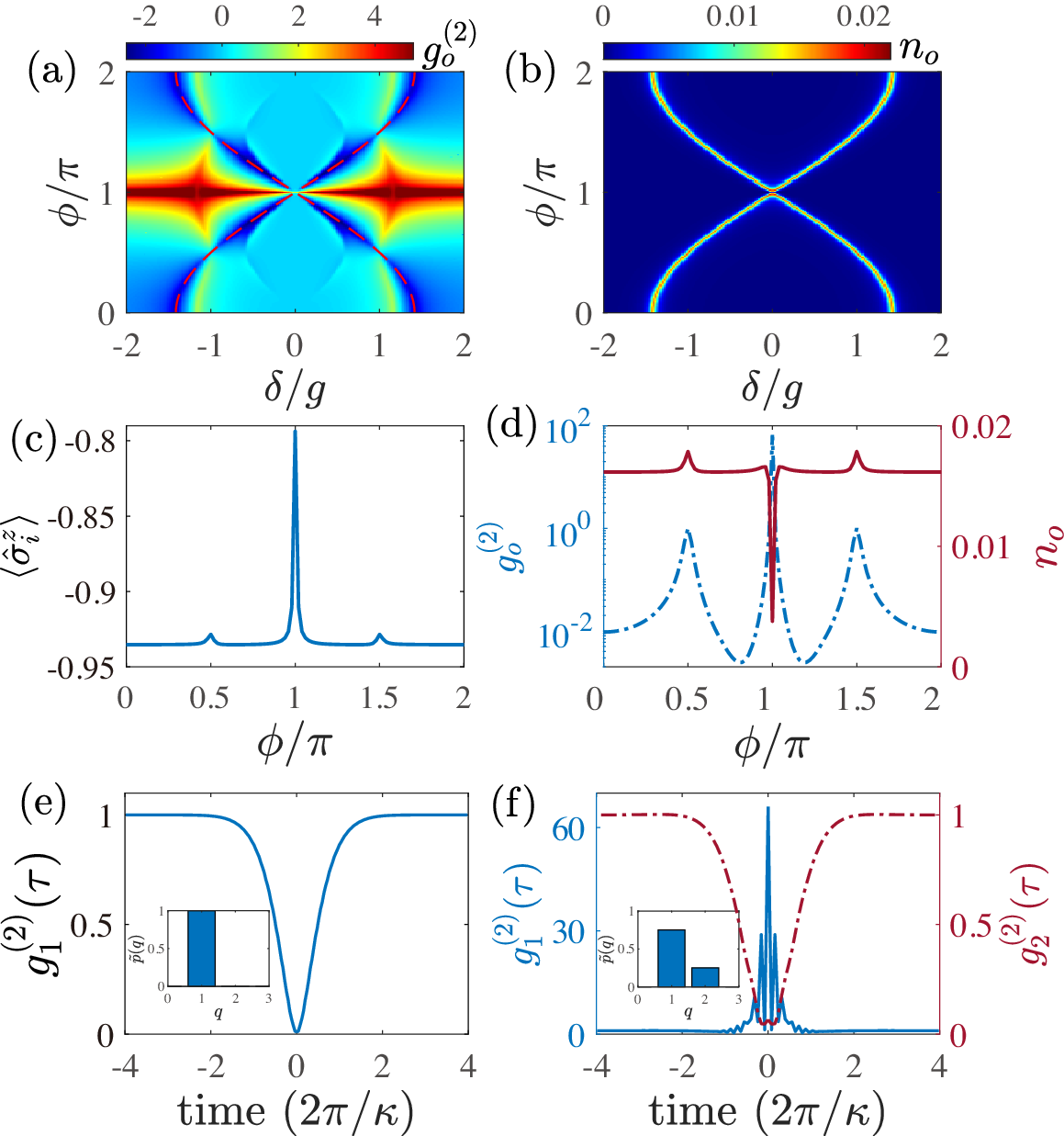}
\caption{Phase-controlled single- to two-photon bundle states for  $\Delta_F=0$. (a) $g_o^{(2)}(0)$ and (b) $n_o$  of cavity mode $\hat{o}$ versus $\delta$ and $\phi$. (c) $\langle \hat{\sigma}^z_i\rangle$ and (d) $g_o^{(2)}$ (dotted-dashed line) and corresponding $n_o$ (solid line) as functions of $\phi$ at $\delta/g = \pm\sqrt{1+\cos\phi}$. (e) $\tau$ dependence of $g^{(2)}$ and $g^{(2)}$ (solid line) and corresponding $g_2^{(2)}$ (dotted-dashed line) at [$\delta/g$, $\phi/\pi$]=[$\sqrt{2}$, 0] and [0, 1], respectively. Inset in (e) and (f) show the distribution $\tilde{p}(q)$ of single- and two-photon states.}
\label{phi}
\end{figure}

The measurable nonreciprocity is captured by master equation including all dissipation channels,
 \begin{equation}\label{master}%
{ \frac{d\rho}{dt}}= -i [\hat{\cal H}, {\rho}] + {\kappa} \mathcal {\cal{D}}[\hat{a}_{\circlearrowright}]\rho  + {\kappa} \mathcal {\cal{D}}[\hat{a}_{\circlearrowleft}]\rho +{\gamma}\sum_{j=1}^2 \mathcal
{\cal{D}}[\hat{\sigma}_{j}^-]\rho,
\end{equation}
where $\rho$ is density matrix of cavity-coupled atomic array, $\mathcal {D}[\hat{o}]\rho=\hat{o} {\rho} \hat{o}^\dag - (\hat{o}^\dag \hat{o}{\rho} + {\rho}\hat{o}^\dag \hat{o})/2$ denotes Lindblad dissipator, and $\kappa$ ($\gamma$) is cavity (atomic) decay rate. To characterize $n$-photon bundle emission, we introduce the generalized correlation function~\cite{C2014Emitters}
\begin{align}
g_n^{(2)}(\tau) = \frac{\left\langle \prod_{i=1}^2 \left[\hat{o}^{\dagger}(\tau_i)\right]^n \prod_{i=1}^2 \left[\hat{o}(\tau_i)\right]^n \right\rangle}{\prod_{i=1}^2 \left\langle \left[\hat{o}^{\dagger}(\tau_i)\right]^n \left[\hat{o}(\tau_i)\right]^n \right\rangle},
\label{g220}
\end{align}
with $\tau=\tau_2-\tau_1$. Genuine $n$-photon bundle emission is identified by $g^{(2)}(0)>g^{(2)}(\tau)$ and $g_n^{(2)}(0)<g_n^{(2)}(\tau)$, indicating intra-bundle bunching and inter-bundle antibunching~\cite{deng2021motional,PhysRevLett.117.203602}. 

For experimental feasibility, we consider two $^{87}$Sr atoms trapped in optical tweezers and coupled to a high-finesse spinning WGM resonator~\cite{PhysRevLett.126.233602,PhysRevLett.134.013403}. We adopt experimentally accessible parameters:  $g=(2\pi)120$ kHz, $\kappa=(2\pi)15$ kHz~\cite{kongkhambut2022observation}, $\gamma=(2\pi)7.5$ kHz, and $\lambda = 689$~nm~\cite{941q-5sdq,PhysRevLett.118.263601}. The weak pump is $\Omega=(2\pi)1.5$ kHz symmetrically addressing CW and CCW modes. The Fizeau splitting is tunable over ${\Delta_F}/g\in[0,1]$, corresponding to slow rotation below $<50\,$Hz. Importantly, the relative atomic separation in ring geometry can be stabilized~\cite{PhysRevLett.134.013403,PhysRevResearch.6.L042026}, rendering fluctuations of $\phi$ negligible, despite sizable single-tweezer position fluctuations~\cite{PRXQuantum.6.010322}. Unlike conventional spinning-resonator schemes  requiring large $\Delta_F$~\cite{fleury2014sound,popa2014non,PhysRevLett.120.125501,PhysRevLett.125.143605}, our mechanism operates at two orders of magnitude smaller shift. Quantum interference strongly amplifies directional response with weak chiral symmetry breaking, thus provides a practical route to nonreciprocal quantum sources~\cite{PhysRevLett.123.233604,PhysRevLett.121.153601,PhysRevLett.133.043601}.

{\em Reciprocal single- to two-photon bundles}.---Before exploring nonreciprocity, we first examine phase-controlled photon emission for nonspinning resonator ($\Delta_F=0$). In this regime, CW and CCW modes are degenerate, yielding reciprocal photon statistics. Figures~\ref{phi}(a) and \ref{phi}(b) show second-order correlation $g_o^{(2)}(0)$ and steady-state photon number $n_{o}$ in $\delta\phi$ plane. The doubly degenerate vacuum-Rabi splitting with symmetric red-blue sidebands evolve continuously with $\phi$, and single-photon resonance occurs at $\delta = \pm g \sqrt{1-\cos\phi}$ [red dashed line in Fig.~\ref{phi}(a)]. Along these resonances, the spin magnetization $\langle \hat{\sigma}^z_i\rangle$ sharply increases at $\phi/\pi=1$ [Fig.~\ref{phi}(c)], while cavity occupation exhibits a pronounced suppression.

Varying $\phi$ drives a continuous crossover between distinct emission regimes [Fig.~\ref{phi}(d)]: strong sub-Poissonian ($g_o^{(2)}=9.5\times10^{-3}$) at $\phi=0$, to Poissonian ($g_o^{(2)}=1$) at $\phi/\pi=0.5$, and super-Poissonian ($g_o^{(2)}=66$) at $\phi/\pi=1$. The suppressed photon occupation coincides with photon bunching at $\phi/\pi=1$, signaling the emergence of nonclassical multiphoton states [Fig.~\ref{phi}(d)]. This behavior originates from destructive interference between excitation pathways~\cite{zhao2025tunable}.  At $\delta=0$ and $\phi=\pi$, the two branches merge to form an atomic dark state within the single-photon manifold, completely suppressing single-photon emission. Such interference-induced dark states are closely related to experimentally observed suppression of cavity emission under direct cavity driving~\cite{PhysRevLett.130.173601,PhysRevLett.131.253603}. A weak transverse drive ($\Omega\neq0$) breaks the perfect blockade of higher manifolds and opens multiphoton excitation channels, thereby generating strongly bunched two-photon bundles~\cite{zhao2025tunable}.

\begin{figure}[ht]
\includegraphics[width=0.95\columnwidth]{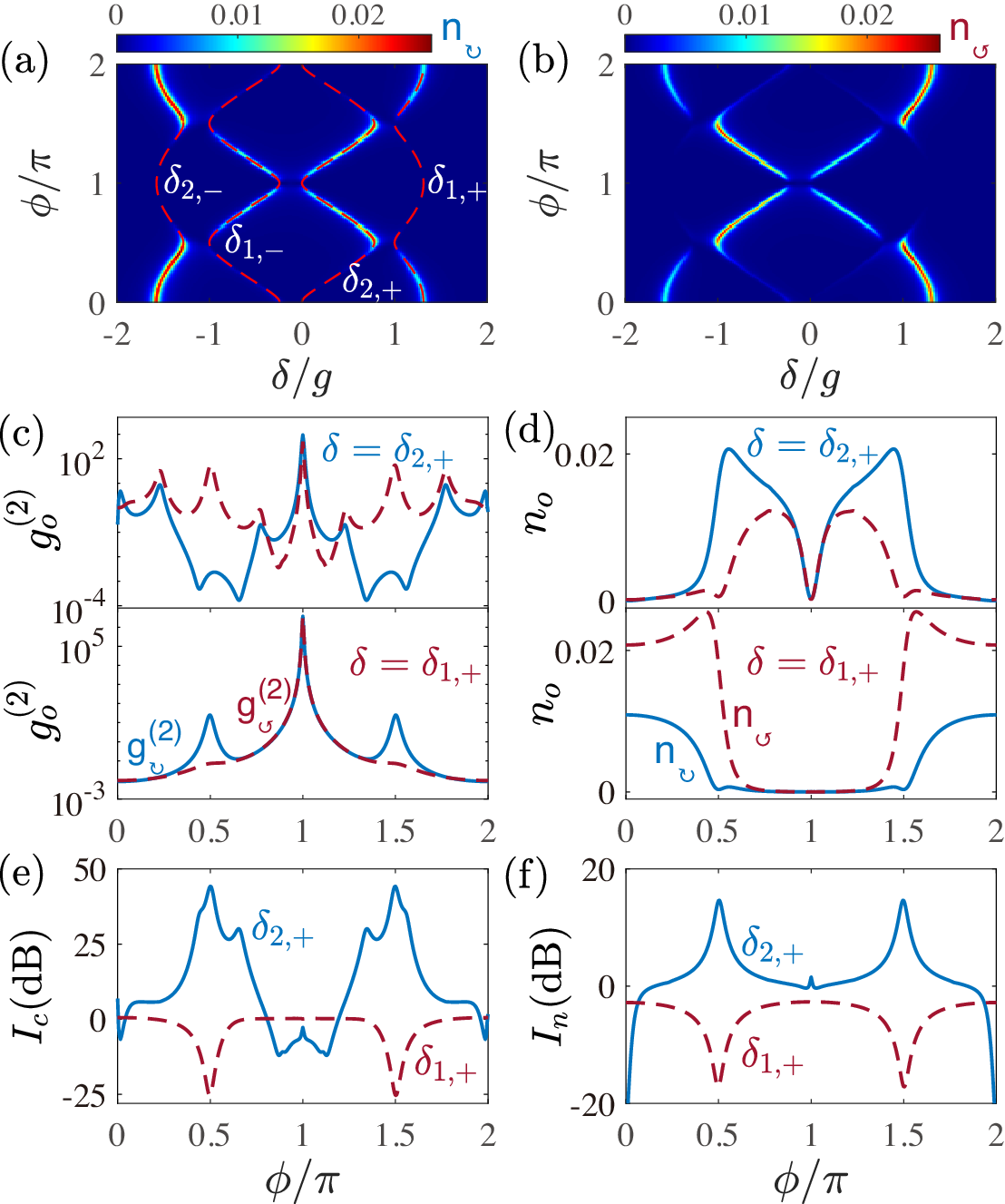}
\caption{Phase-amplified quantum nonreciprocity for spinning resonator at $\Delta_F/g=0.5$.  (a) $n_{\circlearrowright}$ and (d) $n_{\circlearrowleft}$ in the $\delta\phi$ parameter plane. (c) $g_{0}^{(2)}$,  (d) $n_{o}$, (e) $I_c$, and (f) $I_n$ as a function of $\phi$ for different single-photon resonances $\delta$, respectively.}
\label{nonrec}
\end{figure}

Further insight can be gained from time-dependent correlation functions $g_1^{(2)}(\tau)$ and $g_1^{(2)}(\tau)$, which characterize single-photon and two-photon bundle emissions. For $\phi=0$ and sideband $|\delta/g|=\sqrt{2}$, the condition $g_1^{(2)}(\tau)>g_1^{(2)}(0)$ demonstrates strong antibunching [Fig.~\ref{phi}(e)], corresponding to a long photon lifetime $\tau=1/\kappa_a=10.6~\mu$s. This photon blockade is ascribed to enhanced spectral anharmonicity suppressing two-photon excitations. The blockade is further confirmed by steady-state photon-number distribution $\tilde{p}(q)=q p(q)/n_o$ with $p(q)=\langle q|\hat{o}^\dagger\hat{o}|q\rangle$, which measures the fraction of $q$-photon states among total emitted photons. The resulting synergy between collective emission and reinforced blockade enables high brightness while maintaining $\tilde{p}(1)=0.9998$, demonstrating a bright high-purity single-photon source. 

In contrast, two-photon bundles emerge when $g_1^{(2)}(0)>g_1^{(2)}(\tau)$ and $g_2^{(2)}(0)<g_2^{(2)}(\tau)$ [Fig.~\ref{phi}(f)], indicating intra-bundle bunching($g_1^{(2)}(0)=66$) and inter-bundle antibunching ($g_2^{(2)}(0)=0.06$)~\cite{PhysRevLett.117.203602, deng2021motional,C2014Emitters}. The occupation of $q$-photon states with $q>2$ is negligible ($6.4\times 10^{-5}$), and steady-state photon number exceeds the value predicted by Mollow physics by over two orders of magnitude~\cite{C2014Emitters}.  These results identify $\phi$ as a versatile control knob for programmable single- and two-photon bundle states with simultaneously high purity and brightness.
 
{\em Phase-amplified quantum nonreciprocity}.---In stationary resonator ($\Delta_F=0$), phase-controlled interference reshapes anharmonic spectrum, suppressing multiphoton excitation to generate photon blockade or enhancing higher-order processes to produce multiphoton bundles. When $\Delta_F\neq 0$, CW and CCW modes become nondegenerate, breaking chiral symmetry, and strongly amplifying directional quantum nonreciprocity by tuning $\phi$.  Figure~\ref{nonrec}(a) and \ref{nonrec}(b)  show photon occupations of CW and CCW modes at $\Delta_F/g=0.5$. Compared with nonrotating resonator, vacuum splitting evolves from two branches into four branches. The single-photon resonances $\delta_{1,\pm}$ and $\delta_{2,\pm}$ calculated from dressed-state spectrum (red dashed line) agree well with fully numerical solutions of Eq.~(\ref{master}). 

Chiral symmetry breaking induces directional asymmetry in photon brightness across all four branches. While quantum nonreciprocity simultaneously emerges in photon correlations at single-photon resonances. Both correlation and brightness nonreciprocity are highly tunable via $\phi$, as shown in Figs.~\ref{nonrec}(c) and \ref{nonrec}(d). For example, at $\delta=\delta_{2,+}$ and $\phi/\pi=0.66$, CW mode exhibits strong antibunching $g^{(2)}_{{\circlearrowright}}=2\times 10^{-4}$, with high brightness $n_{{\circlearrowright}}=0.18$, whereas CCW mode exhibits weak antibunching $g^{(2)}_{{\circlearrowleft}}=0.16$ with $n_{{\circlearrowleft}}=0.1$. This produces correlation nonreciprocity spanning over three orders of magnitude, even as brightness nonreciprocity remains below a factor of two. Remarkably, anti-bunching and bunching coexist in opposite directions, e.g., $g^{(2)}_{{\circlearrowright}}=0.07$ and $g^{(2)}_{{\circlearrowleft}}=24.7$ at $\phi/\pi=0.5$. Unlike conventional approaches requiring large $\Delta_F$ ($10^1$ MHz) in fast spinning resonator~\cite{fleury2014sound,popa2014non,PhysRevLett.120.125501,PhysRevLett.125.143605,PhysRevLett.120.023601}, our mechanism exploits a weak Fizeau shift ($10^2$ kHz), where phase-controlled interference strongly amplifies directional quantum nonreciprocity.

To quantify directional response, we introduce isolation ratios for photon correlations and brightness,
\begin{align}
I_c(\delta,\Delta_F) &= 10\log({g_{\circlearrowright}^{(2)}}/{g_{\circlearrowleft}^{(2)}}), \nonumber \\
I_n(\delta,\Delta_F) &= 10\log({n_{\circlearrowleft}}/{n_{\circlearrowright}}) ,
\end{align}
which characterize asymmetry between CC and CCW modes. Figures~\ref{nonrec}(e) and \ref{nonrec}(f) show $\phi$ dependence of $I_c$ and $I_n$ at different single-photon resonance. For $\delta=\delta_{2,+}$ and $\phi/\pi=0.5$, correlation isolation reaches $I_c = 45$~dB, accompanied by brightness isolation of $I_n =15$~dB. This demonstrates that strong antibunching can coexist with high emission brightness, contrary to conventional photon-blockade scenarios where strong antibunching typically reduces brightness. Similarly, at $\delta=\delta_{1,+}$ and $\phi/\pi=0.5$, the isolation ratios reach $I_c=-25$ dB and  $I_n=-17$ dB, indicating that CCW mode can operate as a high-quality directional single-photon source. By contrast, for $\phi/\pi = 0$, both correlation and brightness isolation remain weak ($I_c = 0.5$, $I_n = -2.8$) at the same resonance, highlighting the essential role of phase-controlled interference in amplifying weak chiral asymmetry.

\begin{figure}[ht]
\includegraphics[width=0.95\columnwidth]{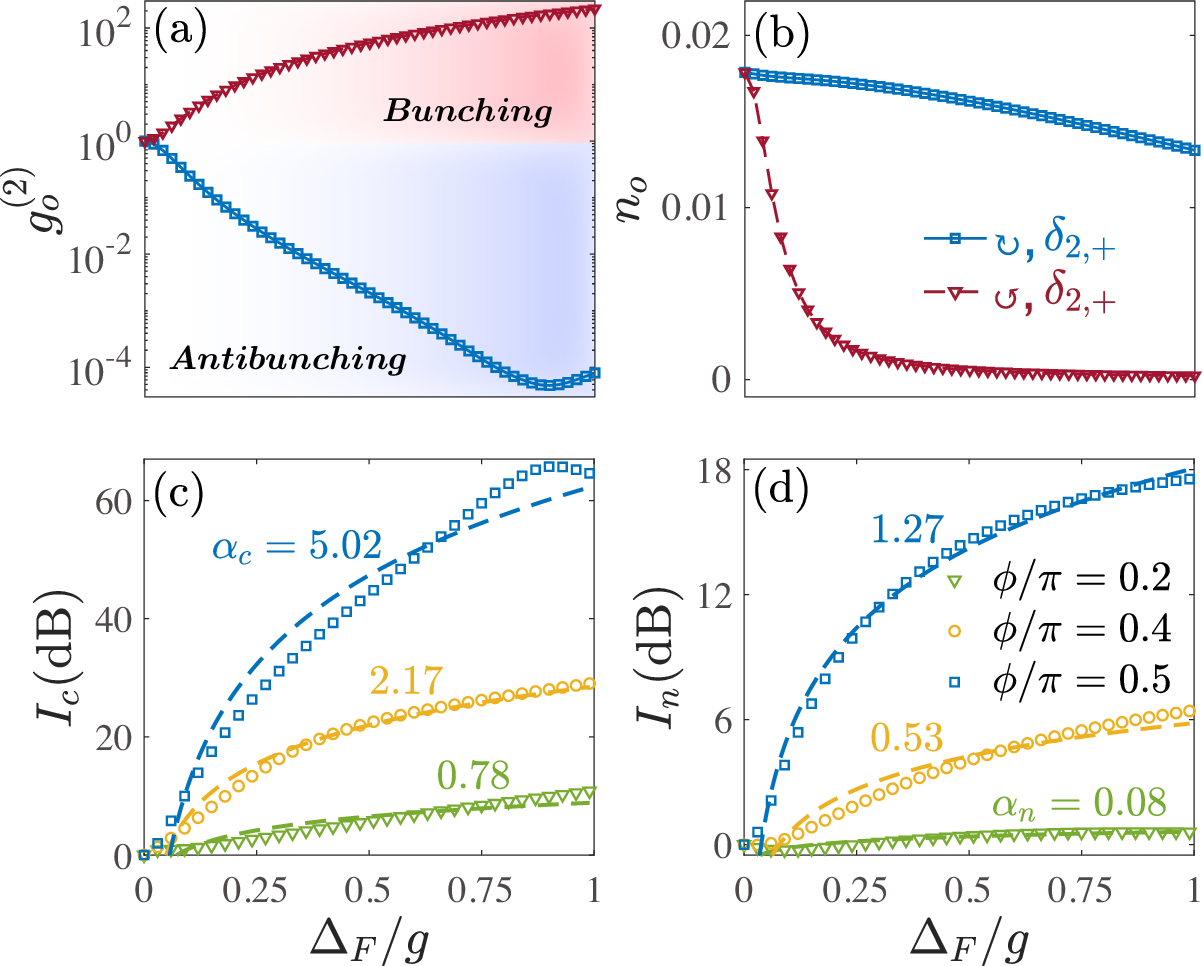}
\caption{Phase-amplified quantum nonreciprocity versus $\Delta_F$ at $\delta=\delta_{2,+}$ and $\phi/\pi=1/2$. (a)-(b) $g_o^{(2)}$ and corresponding to $n_o$ as functions of $\Delta_F$.  (c)-(d) Isolation ratios of $I_c$ and $I_n$ for different $\phi$. Dashed lines show log-log scale fits, with power-law exponents $\alpha_c=[0.78, 2.17, 5.02]$ and $\alpha_n=[0.08,0.53,1.27]$ for $\phi/\pi = [0.2, 0.4, 0.5]$, respectively. 
}
\label{fig4}
\end{figure} 

{\em Power-law enhanced isolation ratio}.---The strong correlation and brightness nonreciprocity can be further amplified via power-law scaling with $\Delta_F$ in the presence of phase-controlled interference. Figures.~\ref{fig4}(a) and~\ref{fig4}(b) show $g^{(2)}_o$ and $n_o$ versus $\Delta_F$ for fixed $\delta=\delta_{2,+}$ and $\phi/\pi=0.5$. At nonspinning point $\Delta_F=0$, CW and CCW modes exhibit reciprocal emission with coherent photon statistics [$g^{(2)}_o=1$] and relatively large intracavity occupation ($n_o=0.018$). This nontrivial behavior indicates that phase-controlled interference strongly suppresses spectral anharmonicity, effectively eliminating photon blockade despite the intrinsic nonlinearity of the system.

For $\Delta_F>0$, chiral symmetry breaking rapidly enhances CW antibunching and CCW bunching. Interestingly, the enhanced CW antibunching occurs with only a slight reduction in brightness. Compared with the reciprocal case at $\Delta_F=0$, the antibunching of CW mode is improved by more than four-orders of magnitude, reaching $g_{\circlearrowright}^{(2)}=4.7\times 10^{-5}$ while maintaining appreciable brightness $n_{\circlearrowright}=0.014$ at $\Delta_F/g=0.9$. This behavior sharply contrasts with conventional photon-blockade schemes~\cite{birnbaum2005photon,dayan2008photon,PhysRevLett.118.133604,PhysRevLett.134.183601,PhysRevLett.134.013602}, where stronger antibunching is typically accompanied by severely reduced output intensity. The coexistence of near-ideal antibunching \emph{without sacrificing brightness} provides a key ingredient for high-quality directional single-photon source. Meanwhile, $g_{\circlearrowleft}^{(2)}$ increases monotonically with $\Delta_F$, indicating the emergence of strongly bunched emission in the opposite direction. Tuning to $\delta=\delta_{1,+}$ reverses quantum nonreciprocity, producing strongly antibunched emission in one direction and weakly antibunched or bunched emission in the other [Fig.~\ref{model}(c)].

To reveal interference-enhanced nonreciprocity, we plot $\Delta_F$ dependence of correlation and brightness isolation ratios for different $\phi$ in Figs.~\ref{fig4}(c) and~\ref{fig4}(d). Both $I_c$ and $I_n$ are strongly amplified with $\phi$ once a finite $\Delta_F$ is introduced. The optimal nonreciprocal response occurs at $\phi/\pi=0.5$, where quantum interference is maximized. In contrast to experimentally reported isolation of $10$ dB near exception points~\cite{song2024experimental}, our scheme already achieves $I_c=15$ dB already in weak Fizeau regime $\Delta_F/g\sim 1$, while brightness isolation increases monotonically with $\Delta_F$. More strikingly, the correlation isolation exhibits a strongly nonlinear dependence on $\Delta_F$, reaching $I_c=65.7$ dB at $\Delta_F/g=0.9$, together with $I_n=17.3$ dB.

The isolation ratios can be fitted by power-law scaling
\begin{equation}
I_{c,n} = A_{c,n}\left({\Delta_F}/{g}\right)^{\alpha_{c,n}},
\end{equation}
where $A_{c,n}$ determines overall isolation magnitude and $\alpha_{c,n}$ is scaling exponent characterizing the sensitivity of nonreciprocity to  Fizeau splitting. Large $\alpha_{c,n}$ indicates stronger amplification of directional asymmetry. Both $\alpha_{c}$ and $\alpha_{n}$ increase rapidly with $\phi$. In particular, the correlation isolation evolves from $\alpha_{c}=0.78$ to $5.02$ as $\phi$ tuned from $0.2\pi$ to $0.5\pi$. Simultaneously, brightness isolation changes from $\alpha_{n}=0.08$ to $1.27$. By contrast, this enhancement weakens significantly away from  $\phi/\pi=0.5$, underscoring the essential role of phase-controlled interference in amplifying both classical and quantum nonreciprocity. The extracted scaling exponents $\alpha_{c,n}$ quantify the nonlinear response of spinning resonator to weak $\Delta_F$. These results establish phase-engineered power-law amplification of weak chirality as a powerful mechanism for chiral quantum technologies~\cite{lodahl2017chiral,bliokh2015quantum}. 

\emph{Conclusion}.--We have proposed a feasible scheme compatible with state-of-the-art cavity-QED platforms, in which phase-controlled quantum interference amplifies weak Fizeau splitting into giant quantum nonreciprocity. By engineering dressed-state spectrum through interference, weak rotation-induced chirality is converted into pronounced directional asymmetry in photon statistics. This mechanism generates a clear separation between antibunched and strongly bunched emission in opposite propagation directions, together with large correlation and brightness isolations. Remarkably, the nonreciprocal response exhibits phase-controlled power-law scaling with Fizeau splitting, enabling giant amplification of weak chiral effects. Our approach is fundamentally distinct from intensity-dependent Kerr-type responses~\cite{shi2015limitations}, reservoir-engineered
nonreciprocal devices~\cite{PhysRevX.5.021025}, and higher-order exceptional points~\cite{hodaei2017enhanced}.

More broadly, our results demonstrate that nonreciprocity can be encoded directly in quantum statistical properties of light, beyond conventional intensity-based nonreciprocal transport. The coexistence of strongly antibunched and bunched emission in opposite directions provides a natural  platform for few-photon nonreciprocal sources~\cite{PhysRevLett.121.153601,PhysRevLett.133.043601,PhysRevLett.123.233604}, where single- and multi-photon states can be spatially separated and selectively addressed. Importantly, interference-assisted amplification relaxes the need for large Sagnac shifts or strong non-Hermitian engineering, allowing strong directional responses from weak chiral symmetry breaking. These results open new opportunities for nonreciprocal quantum sensing~\cite{PhysRevA.103.042418,wang2024quantum,lau2018fundamental}, optical transistor switching~\cite{xiong2024electrical,yang2020inverse}, directional quantum information processing~\cite{PhysRevLett.126.223603,PhysRevLett.134.196904}, and quantum metrology~\cite{dowling2008quantum}.

{\em Acknowledgments}.---This work was supported by the National Natural Science Foundation of China (Grant No.12374365, Grant No. 12274473, and Grant No. 12135018), Quantum Science and Technology-National Science and Technology Major Project (Grant No.2025ZD0300400), Guangdong Provincial Quantum Science Strategic Initiative (Grants No. GDZX2505001), and Guangdong University of Technology SPOE Seed Foundation (SF2024111504).


%

\end{document}